\documentclass{llncs}

\usepackage[dvips]{graphicx}
\usepackage{amsmath}
\usepackage{amsfonts}
\usepackage{latexsym}
\usepackage{amssymb}
\usepackage{epsfig}
\usepackage[mathcal]{euscript}

\newcommand{\VL}[1]{#1}
\newcommand{\VC}[1]{}

\newcommand{\couic}[1]{}

\newcommand{\ovb}[1]{\overline{#1}}
\newcommand{\vt}{V}
\newcommand{\dom}{\textrm{dom}}
\newcommand{\ports}{1..\pi}
\newcommand{\port}{\!:\!}

\def\N{\mathbb{N}}

\begin{document}
\title{Causal graph dynamics}
\author{Pablo Arrighi\inst{1,2} \and Gilles Dowek\inst{3}}
\institute{
Universit\'e de Grenoble, LIG, 220 rue de la chimie, 38400 SMH, France\\
\and
\'Ecole Normale Sup\'erieure de Lyon, LIP, 46 all\'ee d'Italie, 69008 Lyon, France
\email{parrighi@imag.fr}
\and
INRIA, 23 avenue d'Italie, CS 81321, 75214 Paris Cedex 13, France.\\
\email{gilles.dowek@inria.fr}}
\maketitle

\begin{abstract}
We\VC{\footnotetext{The full version of this paper is available as arXiv:1202.1098.}} extend the theory of Cellular Automata to arbitrary, time-varying graphs.
\VL{In other words we formalize, and prove theorems about, the intuitive idea of a labelled graph which evolves in time --- but under the natural constraint that information can only ever be transmitted at a bounded speed, with respect to the distance given by the graph. The notion of translation-invariance is also generalized. The definition we provide for these `causal graph dynamics' is simple and axiomatic. The theorems we provide also show that it is robust. For instance, causal graph dynamics are stable under composition and under restriction to radius one. In the finite case some fundamental facts of Cellular Automata theory carry through: 
causal graph dynamics admit a characterization as continuous functions, 
and they are stable under inversion. The provided examples suggest a wide range of applications of this mathematical object, from complex systems science to theoretical physics.\\}
\VL{{\bf Keywords.} {\em Dynamical networks, Boolean networks, Generative networks automata, Cayley cellular automata, Graph Automata, Graph rewriting automata, Parallel graph transformations, Amalgamated graph transformations, Time-varying graphs, Regge calculus, Local, No-signalling.}}
\end{abstract}

\section{Introduction}

\noindent {\em A question.} There are countless situations in which some agents (e.g. physical systems \cite{MeyerLove}, computer processes \cite{PapazianRemila}, biochemical agents \cite{MurrayDicksonVol2}, economical agents \cite{KozmaBarrat}, users of social networks, etc.) interact with their neighbours, leading to a global dynamics, the state of each agent evolving through the interactions. In most of these situations, the topology, i.e. who is next to whom also varies in time (e.g. agents become physically connected, get to exchange contact details, move around, etc.). The general concept of a dynamics caused by neighbour-to-neighbour interactions and with a time-varying neighbourhood, is therefore quite natural.\\
At the mathematical level, however, this general concept turns out to be rather difficult to formalize. \VL{We can, of course, readily define some specific models which appear to fall in this category. But the more foundational question of understanding what defines `causal dynamics with time-varying neighbourhoods' in general, and what fundamental facts arise from the rigorous definitions we may put forward, is a slippery one.}There are at least three difficulties. The first is that the neighbourhood relation plays a double role in this story, as it is both a constraint upon the global dynamics, and a subject of the global dynamics, which modifies it. The second is that, as agents get created and deleted, the notion of who has become whom is not so obvious, but this notion is needed in order to state the causality property that only neighbours may communicate in one step of time. The third is to express that the global dynamics should `act everywhere the same', a property akin to translation-invariance\ldots but arbitrary graphs do not admit such translations.

\VL{\noindent {\em Two scenarios.} In a mobile phone network, the agents could be the mobile phones, and their neighbours the mobile phones that they can call, i.e. the ones whose phone numbers is in their list of contacts. In this picture, the global state of the network can then thought of as a directed labelled graph, with each vertex representing a mobile phone having some internal state, and the edges representing the neighbourhood relation. The entire graph evolves in time, but this global dynamics emerges from a neighbour-to-neighbour interactions. In other words, during one call duration, information can only propagate to the nearest neighbour. Although neighbours change in time, they do in a relatively controlled manner: new contacts are always contacts of contacts. Sometimes new mobile phones get created, and others get thrown out.\\
In general relativity, the agents could be some particles which lie at particular locations of a smooth Riemannian manifold, and the neighbours those particles which lie not too far away in the manifold, due to the absolute bound on the speed of information. The notion of bounded speed, however, is defined relative to the metric of the manifold, which changes in time according to the mass of the particles. Again it changes in a relatively controlled manner, as far-away perturbations in mass repartition should not impact the local metric. Sometimes particles scatter and others get emitted.}

\noindent {\em Two approaches.} Cellular Automata research lies at the cross-point between Physics, Mathematics, and Computer Science. Cellular Automata consist of a grid of identical square cells, each of which may take one of a finite number of possible states. The entire array evolves in discrete time steps. The dynamics is required to be translation-invariant (it commutes with translations of the grid) and causal (information cannot be transmitted faster than a fixed number of cells per time step). Whilst Cellular Automata are usually defined as exactly the functions having those physics-like symmetries, it turns out that they can also be characterized in purely mathematical terms as the set of translation-invariant continuous functions \cite{Hedlund} for a certain metric. Moreover in a more Computer Science oriented-view, Cellular Automata can be seen as resulting from the synchronous application of the same local rule throughout the grid.
These three complementary ways (physical causality, mathematical continuity and constructive locality) of addressing Cellular Automata are what anchors this field in rigorous foundations.\VL{This has led a large body of theoretical (set theoretical properties, dynamical properties, decidability properties, connections with tilings etc.) and applicative works (modelling of anything from sea shells to traffic jams).}
Still, restricting to a fixed grid has been perceived to be a limitation. As a consequence Cellular Automata definitions have been extended from grids to Cayley or hyperbolic graphs, where most of the theory carries though \cite{Roka,Coornaert,Margenstern}. But these graphs are quite regular in many respects, for instance they are self-similar under translations. More recently Cellular Automata definitions have been extended to graphs \cite{PapazianRemila,Gruner}, in order to describe certain distributed algorithms. In these extensions, the topology remained fixed: they would be unable to account for evolving mobile networks, nor discrete formulations of general relativity \cite{Sorkin}. {\em This paper aims at extending Cellular Automata theory to arbitrary, time-varying graphs. The theorems we prove are mainly inspired by those at the foundation of Cellular Automata theory. Two of them show the equivalence of causality with local rule constructions (Theorem \ref{th:structure}) and continuity (Theorem \ref{th:continuity}), and two others are closure properties by composition (Theorem \ref{th:composability}) and inversion (Theorem \ref{th:reversibility}).}

The second, related line of work is that of Graph Rewriting. The idea of rewriting graphs by applying some replacement rules has arisen as a natural generalization of term rewriting, and is now widespread in Computer Science \cite{RozenbergBook,EhrigBook}. \VL{For instance it has been used in order to model situations ranging from the representation of programs with pointers and memory safety properties, to Object-oriented software engineering paradigms.} Whilst Cellular Automata theory focuses on changing states\VL{ and not the topology}, Graph Rewriting focuses on changing the topology\VL{ and not the states}. But there are other fundamental differences. Whilst Cellular Automata theory focuses on the global dynamics resulting from the synchronous application of the same local rule throughout the graph, Graph Rewriting theory usually focuses on asynchronous applications of a rule which need not be local, this leading to an undefined global dynamics (hence the emphasis on properties such as confluence, non-interference, etc.). Amalgamated Graph Transformations \cite{BFHAmalgamation,LoweAlgebraic} and Parallel Graph Transformations \cite{EhrigLowe,Taentzer,TaentzerHL} are noticeable exceptions in this respect, as they work out rigorous ways to apply local rewriting rules synchronously throughout the graph. Still the properties of the resulting global dynamics are not examined. {\em This paper aims at extending the focus of Graph Rewriting to changing states, as well as to deduce aspects of Amalgamated/Parallel Graph Transformations from the axiomatic properties of the global dynamics.}

\noindent {\em Third way.} The idea of a rigorous model of computation in which both the states and the topology evolve is certainly not new, and can be attributed to Kolmogorov and Upsenskii \cite{KolmogorovUspensky}, see also \cite{Schonhage}. These models, as well as the more recent \cite{SamayaGNA,Cavaliere}, are again asynchronous in a sense. There is no spatial parallelism, although it may be simulated \cite{TomitaGRAAsync}. Lately, several groups have taken a more practical approach to this problem, and have started to develop simulation environments \cite{GiavittoMGS,Mammen,Kurth} based on different programming paradigms, all of them implementing the idea of rewriting both the graph and its states via repeated applications of some replacement rules. These systems offer the possibility to apply the replacement rules simultaneously in different non-conflicting places. Such an evaluation strategy does provide some weak form of synchronism. Sometimes, when a set of rule is non-conflicting, this evaluation strategy happens to coincide with full synchronism. This pragmatic approach to extending Cellular Automata to time-varying graphs is advocated in \cite{TomitaGRA,KreowskiKuske,MeyerLove}, and has led to some advanced algorithmic constructions \cite{TomitaSelfReproduction,TomitaSelfDescription}. {\em This paper aims at proposing simple formalizations of the notions of translation-invariance and causality when referring to functions from labelled graphs to labelled graphs. It aims at deducing, from these two notions, what is the most general form of fully synchronous application of a local rule throughout a graph, as well as to achieve a full characterization of what are the suitable local rules.}

\VL{\noindent {\em This paper.} Section \ref{sec:graphdynamics} provides an equivalent of the translation-invariance condition of Cellular Automata dynamics; since general graphs do not admit translations this becomes an isomorphism-invariance. Section \ref{sec:causalgraphdynamics} provides the axiomatic definition of causal graph dynamics, and shows its equivalence with a more constructive definition expressed in terms of synchronous applications of local rules. It also provides two canonical examples of causal graph dynamics: Cellular Automata, the Inflating grid. Section \ref{sec:stability} provides general robustness theorems such as stability under composition and restriction to radius one. Section \ref{sec:continuity} relates causality and continuity. As in the theory of Cellular Automata, we obtain that causal graph dynamics admit a characterization as continuous functions. In Section \ref{sec:invertibility} we also show that they are stable under inversion. Section \ref{sec:conclusion} provides a summary of the main results of this paper, also useful as a reading guide. It mentions numerous tracks for future research.}

\section{Graphs dynamics}\label{sec:graphdynamics}

\VL{\subsection{Graphs}\label{subsec:graphs}}

\VC{\noindent {\em Graphs.}} We fix an uncountable infinite set $V$ of names. 
The {\em vertices} of the graphs we consider in this paper are uniquely identified by a name $u$ in $V$. 
Vertices may also have a {\em state} $\sigma(u)$ in $\Sigma$.
Each vertex has several {\em ports}, numbered between $1$ and 
a natural number $\pi$. A vertex and its port are written $u \port i$.  
An {\em edge} is a pair $(u \port i, v \port j)$. 
Edges also have a {\em state} $\delta(u \port i, v \port j)$ in $\Delta$. 
In fact, the presence of an edge is held in the domain of the function $\delta$.
\begin{definition}[Graph]
A {\em graph} $G$ with states $\Sigma,\Delta$ and degree $\pi$ is given by 
\begin{itemize}
\item An at most countable subset $\vt(G)$ of $V$ whose elements are called {\em vertices}.
\item A set $\ports$ whose elements are called {\em ports}.
\item A partial function $\sigma$ from $\vt(G)$ to $\Sigma$ giving the state of the vertices.
\item A partial function $\delta$ from $(\vt(G)\port\ports)\times(\vt(G)\port\ports)$ to $\Delta$ giving the states of the edges, such that each $u\port i$ in $(\vt(G)\port\ports)$ appears at most once in $\dom(\delta)$.
\end{itemize}
The {\em set of all graphs} with states $\Sigma,\Delta$ and degree $\pi$ is written ${\cal G}_{\Sigma,\Delta,\pi}$. 
To ease notations, we sometimes write $v \in G$ for $v \in \vt(G)$.
\end{definition}
The definition is tailored so that we are able to cut the graph around some vertices, whilst keeping the information about the connectivity with the surrounding vertices. The choice of $V$ uncountable but $\vt(G)$ countable allows us to always pick fresh names. The edges are oriented, but for our dynamics to be compositional we need to define neighbours regardless of edge orientation:
\begin{definition}[Neighbours]
We write $u\frown v$ if there exists ports 
$i,j\in\ports$ such that either $\delta(u\port i,v\port j)$ or 
$\delta(v\port j,u\port i)$ is defined.
We write $u\frown^k v$ if there exists $w_1,\ldots,w_{k-1}$ such that $u\frown w_1\frown\ldots\frown w_{k-1}\frown v$.
We write $u\frown^{\leq r} v$ if there exists $k\leq r$ such that $u\frown^k v$.
The {\em set of neighbours of radius $r$ (a.k.a. of diameter $2r+1$)} of a set $A$ with respect to a graph $G$ is the set of the vertices $v$ in $\vt(G)$ such that $u\frown^{\leq r} v$, for $u$ in $A$.
\end{definition}
Moving on, a {\em pointed graph} is just a graph where one, or sometimes several, vertices are privileged. 
\begin{definition}[Pointed graph]
A {\em  pointer set} of  $G$ is a  subset of $\vt(G)$. A  {\em pointed
graph} is given by a graph $G$ and pointer set $A$ of $G$.  The {\em
set of pointed graphs}  with states $\Sigma,\Delta$ and degree $\pi$
is written ${\cal P}_{\Sigma,\Delta,\pi}$.
\end{definition}

\VL{\subsection{Dynamics}\label{subsec:graphsdynamics}}

\VC{\noindent {\em Dynamics.}} Consider an arbitrary function from graphs to graphs. The difficulty we now address is that of expressing the condition that this function `acts everywhere the same' --- a property similar to that of translation-invariance in the realm of Cellular Automata. Of course arbitrary graphs do not admit translations; the first idea is that translation-invariance becomes an invariance under {\em isomorphisms}. In other words, the names of the vertices are somewhat immaterial, they can be renamed, unlike states and ports.
\begin{definition}[Isomorphism]
An {\em isomorphism} $R$ is a function from ${\cal G}_{\Sigma, \Delta, \pi}$ to ${\cal G}_{\Sigma, \Delta, \pi}$ which is specified by a bijection $R(.)$ from $V$ to $V$. 
The image of a graph $G$ under the isomorphism $R$ is a graph $RG$ whose set of vertices is $R(\vt(G))$, and whose partial functions $\sigma_{RG}$ and $\delta_{RG}$ are the compositions $\sigma_G\circ {R}^{-1}$ and $\delta_G\circ {R}^{-1}$ respectively. When $G$ and $H$ are isomorphic we write $G\approx H$. 
Similarly, the image of a pointed graphs $P=(G,A)$ is the pointed graph $RP=(RG,R(A))$. When $P$ and $Q$ are isomorphic we write $P\approx Q$. 
\end{definition}

It would seem that the graph {\em dynamics} we are interested in must commute with isomorphisms, as in \cite{Gabbay}. Unfortunately, demanding straightforward commutation with isomorphisms would enter in conflict with the possibility to introduce new vertices.
\begin{proposition}[Commuting forbids new vertices]\label{lem:commutingimpliesnonincreasing}
Let $F$ be a function from ${\cal G}_{\Sigma, \Delta, \pi}$ to ${\cal G}_{\Sigma, \Delta, \pi}$, which {\em commutes with isomorphisms}, i.e. such that for any isomorphism $R$, $F \circ R = R \circ F$. Then for any graph $G$, $\vt(F(G))\subseteq \vt(G)$.
\end{proposition}

\VL{\proof{
Say there exists $v\in F(G)$ but $v\notin G$. Then we can take some $R$ such that $R(v)$ is neither in $G$ nor in $F(G)$, and $RG=G$. We then have $R(v)\in \vt(RF(G))=\vt(F(RG))=\vt(F(G))$, a contradiction.\hfill $\Box$}}

\noindent The question of circumventing this limitation has been pointed out to be a difficulty \cite{Sieg}. We propose to focus on the following, weaker property instead:
\begin{definition}[Dynamics]\label{def:dynamics}
A {\em dynamics} $F$ is a function from ${\cal G}_{\Sigma, \Delta, \pi}$ to ${\cal G}_{\Sigma, \Delta, \pi}$, such that the following two conditions are met:
\begin{itemize}
\item[(i)] {\em Conjugacy.} For any isomorphism $R$ there exists an
isomorphism $R'$, called a {\em conjugate} of $R$ through $F$, such that 
$F \circ R = R' \circ F$, i.e. for all $G$, $F(RG)=R'F(G)$.
\item[(ii)] {\em Freshness.} For any family of graphs $(G^{(i)})$, $\big[\;\bigcap \vt(G^{(i)}) =\emptyset\;\Rightarrow\;\bigcap \vt(F(G^{(i)}))=\emptyset\;\big]$.
\end{itemize}
The definition extends to functions from ${\cal P}_{\Sigma, \Delta, \pi}$ to ${\cal G}_{\Sigma, \Delta, \pi}$ in the obvious way.
\end{definition}

Note that both conditions in the above definition would have been
entailed by straightforward commutation of $F$ with every
$R$. Moreover, condition $(ii)$ still has a natural interpretation:
to generate a fresh common vertex (inside $\bigcap \vt(F(G^{(i)}))$)
two parties (inside the $(G^{(i)})$) must use a common resource (inside $\bigcap \vt(G^{(i)})$). Even two persons bumping into each other in the street share a common medium: the street.
Note that, due to this condition, dynamics send the empty
graph to the empty graph.

\VL{\section{Causal and localizable dynamics}\label{sec:causalgraphdynamics}}
\VL{\subsection{Causal dynamics}\label{subsec:causalgraphdynamics}}
\VC{\section{Causal dynamics}\label{subsec:causalgraphdynamics}}

Our goal in this \VL{Subsection}\VC{Section} is 
to define the notion of causality in an
axiomatic way. Intuitively a dynamics is {\em causal} if the state and connectivity
of each image vertex is determined by a small graph 
describing the state and connectivity of the neighbours of some 
antecedent vertex. Moreover, not only each 
image vertex is generated by an antecedent vertex, but also each antecedent vertex
generates only a bounded number of image vertices. 
In order to make this into a formal definition we clearly need a notion of {\em disk} around a set of vertices, as well as a notion of {\em antecedent}.

\begin{definition}[Induced subgraph]
The {\em induced subgraph} of a graph $G$ around a set $U$ is a graph $G_U$:
\begin{itemize}
\item whose vertices $\vt(G_U)$ are given by the neighbours of radius one of the set $U'=(\vt(G)\cap U)$.
\item whose partial function $\sigma_{G_U}$ is the restriction of $\sigma$ to $U'$. 
\item whose partial function $\delta_{G_U}$ is the restriction of $\delta$ to 
$$\big((\vt(G_U)\port\ports)\times (U'\port\ports)\big)\cup \big((U'\port\ports)\times(\vt(G_U)\port\ports)\big).$$
\end{itemize}
\end{definition}

\begin{definition}[Disk]
The {\em envelope of radius $r$ (a.k.a. of diameter $2r+1$)} of a set $A$ with respect to a graph $G$ is the pointed graph whose graph is the induced subgraph of $G$ around the neighbours of radius $r$ of $A$, and whose pointer set is $A$ itself. It is denoted $G^r_A$. 

The {\em set of disks of radius $r$}
(a.k.a. of diameter $2r+1$)  
of the set of graph ${\cal G}_{\Sigma, \Delta, \pi}$ 
is the set of envelopes $\{G^r_v\,|\,G\in {\cal G}_{\Sigma, \Delta, \pi}\}$, 
i.e. those centered on a single vertex. 
It is denoted ${\cal D}^r_{\Sigma, \Delta, \pi}$. 
The {\em set of all disks} 
of the set of graph ${\cal G}_{\Sigma, \Delta, \pi}$ is $\bigcup_r {\cal D}^r_{\Sigma, \Delta, \pi}$.
It is denoted ${\cal D}_{\Sigma, \Delta, \pi}$.
\end{definition}

\begin{center}
\epsfig{file=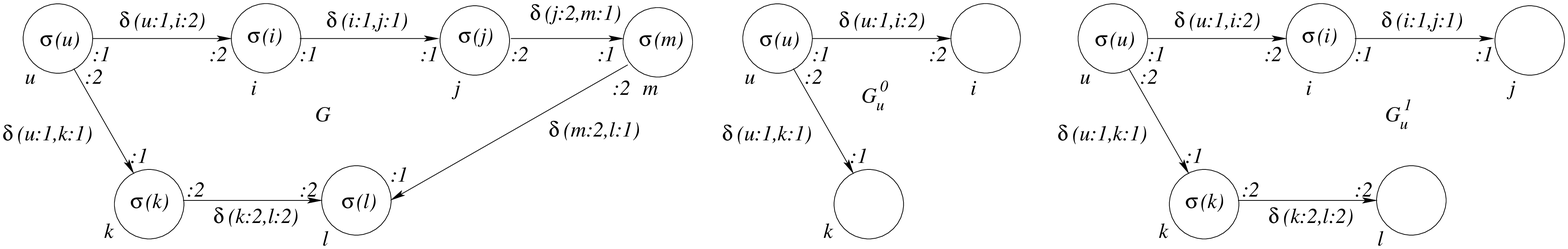,height=2cm}
\end{center}

Dynamics make vertex names somewhat immaterial, but still do not prevent from using these names in order to identify the antecedent vertices of an image vertex.
\begin{definition}[Antecedent codynamics]
Let $F$ be a dynamics from ${\cal G}_{\Sigma, \Delta, \pi}$ to ${\cal G}_{\Sigma, \Delta, \pi}$. We define the antecedent codynamics $a()$ from $V$ to subsets of $V$ such that $v\in a(v')$ if and only if:
$$\forall G, [v'\in F(G) \Rightarrow v\in G].$$
\end{definition}

\VL{\begin{lemma}[Properties of antecedents codynamics]
\label{lem:antecedentcodynamics}
$$\forall G,\forall v'\in F(G),~ a(v')\subseteq G$$
$$\forall v',~ |a(v')|\geq 1$$
$$\forall R,~ \big[R\circ a=a \circ R'\big]$$
with $R'$ a conjugate of $R$ through $F$. The word codynamics refers to this property.
\end{lemma}}

\VL{\proof{$[\textrm{Generacy}]$ Say $v\in a(v')$. We have that $\forall H$,
$[v'\in F(H)\Rightarrow v\in H]$. In particular $v'\in F(G)$ implies $v\in G$.\\
\noindent  $[\textrm{Non-emptiness}]$ 
\begin{align*}
v\in a(v') &\Leftrightarrow \forall G,\, [v'\in F(G) \Rightarrow v\in G]\\
&\Leftrightarrow v\in\bigcap_{v'\in F(G)} G
\end{align*}
since $\bigcap_{v'\in F(G)} \vt(F(G))$  is non-empty, $\bigcap_{v'\in F(G)} \vt(G)$ is non-empty by the freshness condition.\\
\noindent $[\textrm{Codynamicity}]$
\begin{align*}
v\in R(a(v')) &\Leftrightarrow R^{-1}v\in a(v')\\
	&\Leftrightarrow \forall G,\, [v'\in F(G) \Rightarrow R^{-1}(v)\in G]\\
	&\Leftrightarrow \forall G,\, [R'(v')\in R'F(G) \Rightarrow v\in RG]\\
	&\Leftrightarrow \forall G,\, [R'(v')\in F(RG) \Rightarrow v\in RG]\\ 
	&\Leftrightarrow \forall H,\, [R'(v')\in F(H) \Rightarrow v\in H]\\ 
	&\Leftrightarrow v\in a(R'(v))
\end{align*}
\hfill$\Box$}}

\noindent The following is our main definition.
\begin{definition}[Causality]\label{def:causal}
A dynamics $F$ from ${\cal G}_{\Sigma, \Delta, \pi}$ to ${\cal G}_{\Sigma, \Delta, \pi}$ is {\em causal} if and only if there exists a radius $r$ and a bound $b$, such that the following two conditions are met:
\begin{itemize}
\item[(i)] Uniform continuity.
$$\forall v',v\in a(v'),\forall G,H,\; \big[G^r_{v}=H^r_{v}\Rightarrow F(G)_{v'}=F(H)_{v'}\,\big]$$
\item[(ii)]  Boundedness.
$$\forall G,\forall v\in G,\; |\{v'\in F(G)\,|\,v \in a(v')\}|\leq b.$$
\end{itemize}
with $a(.)$ the antecedent codynamics of $F$.
\end{definition}
This definition captures, in a formal way, the physical idea that information propagate at a bounded velocity.

\VL{\subsection{Localizable dynamics}}
\VC{\section{Localizable dynamics}}

The definition of causality does not provide us with a concrete way to construct such causal dynamics. 
We now introduce the more concrete notion of {\em localizable dynamics}, i.e. a dynamics which is induced by a local rule.
\VL{The relationship between these two notions is postponed till Subsection \ref{subsec:causalislocal}.}
Hence let us thus shift our focus towards bottom-up, local mechanisms for rewriting graphs in a parallel manner.
This construction is reminiscent of \cite{BFHAmalgamation,LoweAlgebraic}. 
\VC{We define the notion of {\em consistency} between two graphs as the fact that they do not disagree on their intersections. The {\em union} of two consistent graphs is itself a graph. See the long version of the paper for more details.}
\VL{
We need a notion of {\em union} of graphs, and for this purpose we need a notion of {\em consistency} between the operands.
\begin{definition} [Consistent]
Consider two graphs $G$ and $H$ in ${\cal G}_{\Sigma,\Delta,\pi}$, they are {\em consistent} if and only if:
\begin{itemize}
\item over the set $U'=(\vt(G)\cap \vt(H))$ the partial functions $\sigma_{G}$ and $\sigma_{H}$ agree when they are both defined, meaning that: $\forall u\in U'$,
$$\big[\sigma_G(u)=s\wedge\sigma_H(u)=s' \Rightarrow s=s'\big]$$
\item over the set $U'$ the partial functions $\delta_{G}$ and $\delta_{H}$ agree when they are both defined, meaning that: $\forall u\port i\in (V'\port\ports),\,\forall v\port j,\,v'\port j'\in\big((\vt(G)\cup\vt(H))\port\ports\big),$
$$\big[\big(\delta_G(u\port i,v\port j)=d\wedge\delta_H(u\port i,v'\port j')=d'\big) \Rightarrow (v=v'\wedge j=j'\wedge d=d')\big]\;\wedge$$
$$\big[\big(\delta_G(v\port j,u\port i)=d\wedge\delta_H(v'\port j',u\port i)=d'\big) \Rightarrow (v=v'\wedge j=j'\wedge d=d')\big]\phantom{\;\wedge}$$
\end{itemize}
The definition extends to pointed graphs by ignoring the pointer sets.
\end{definition}

\begin{definition} [Union]
Consider two consistent graphs $G$ and $H$, we define the graph
$G\cup H$ to be the graph:
\begin{itemize}
\item whose set of vertices $\vt(G\cup H)$ is $\vt(G)\cup \vt(H)$. 
\item whose partial function $\sigma_{G\cup H}$ has domain $\dom(\sigma_{G})\cup\dom(\sigma_{H})$ and coincides with $\sigma_{G}$ (resp. $\sigma_H$) over $\dom(\sigma_G)$ (resp. $\dom(\sigma_H)$).
\item whose partial function $\delta_{G\cup H}$ has domain $\dom(\delta_{G})\cup\dom(\delta_{H})$ and coincides with $\delta_G$ (resp. $\delta_H$) over $\dom(\delta_G)$ (resp. $\dom(\delta_H)$).
\end{itemize}
The definition extends to pointed graphs by making the union of the pointer sets.
\end{definition}
}
Roughly speaking the result of a localizable dynamics is the union of the result of the parallel application of a {\em local rule}, \VL{a concept} which we now define.
\begin{definition}[Consistent function]
A function $f$ from ${\cal D}^r_{\Sigma, \Delta, \pi}$ to 
${\cal G}_{\Sigma, \Delta, \pi}$ is {\em consistent} if and only if for any graph $G$, for any pair of disks $G^r_u$, $G^r_v$, $f(G^r_u)$ is consistent with $f(G^r_v)$.
\end{definition}
\begin{definition}[Bounded function]
A subset of graphs ${\cal G}_{\Sigma, \Delta, \pi}$ is {\em bounded} if there exists $b$ such that for any graph $G$ in the set, $|\vt(G)|$ is less or equal to $b$. A partial function from ${\cal G}_{\Sigma, \Delta, \pi}$, or ${\cal P}_{\Sigma, \Delta, \pi}$, to ${\cal G}_{\Sigma, \Delta, \pi}$ is  {\em bounded} if its co-domain is bounded. 
\end{definition}
\begin{definition}[Local rule]
A function from ${\cal D}^r_{\Sigma, \Delta, \pi}$ to ${\cal
G}_{\Sigma, \Delta, \pi}$ is a {\em local rule or radius $r$} if it is a consistent
bounded dynamics.
\end{definition}

Here is the natural way to parallelize the application of the above-described local rules into a global dynamics.
\begin{definition}[localizable dynamics]
A dynamics $F$ from ${\cal G}_{\Sigma, \Delta, \pi}$ to ${\cal G}_{\Sigma, \Delta, \pi}$ is {\em localizable} if and only if there exists $r$ a radius and $f$ a local rule from ${\cal D}^r_{\Sigma, \Delta, \pi}$ to ${\cal G}_{\Sigma, \Delta, \pi}$ such that for every graph $G$ in ${\cal G}_{\Sigma,r}$, 
$$F(G)=\bigcup_{v\in G} f(G^r_v).$$
\end{definition}

\VL{\subsection{Examples of localizable dynamics}}

There are many, more specific-purpose models which a posteriori can be viewed as instances of the notion of causal graph dynamics developed in this paper \cite{MeyerLove,PapazianRemila,Lathrop,MurrayDicksonVol2}. We now show via two examples that our model subsumes, but is not restricted to, Cellular Automata.

\noindent {\em Cellular automata}. For some localizable dynamics, the graph is finite and does not change. Such dynamics are called {\em bounded cellular automata}. Some others slightly expand their border.  Such
dynamics are called {\em finite unbounded cellular automata}. One-dimensional finite unbounded cellular automata are usually defined as follows.

\begin{definition}[Finite unbounded cellular automata]
An alphabet $\Sigma$ is a finite set of symbols, with a distinguished
quiescent symbol $q$.  A configuration $c$ is a function $\mathbb{Z}
\rightarrow \Sigma$, that is equal to $q$ almost everywhere. A {\em
  finite unbounded cellular automaton} is a function $H$ from
configurations to configurations such that
$H(c)_{i+1}=h(c_i,c_{i+1})$, with $h$ a function from $\Sigma^2$ to
$\Sigma$ such that $h(q,q)=q.$
\end{definition}
Note that the use of quiescent states is an artifact to
express that configurations are finite but unbounded,
i.e. that they may grow arbitrarily large. 

A configuration $c$ of such a Cellular Automaton can be represented 
as a finite graph as
follows: there is an interval $I =
[n,p]$ such that $c_i=q$ whenever $i \notin I$. We take
$\vt(G) = I$, $\pi = 2$, $\delta(x\port
2,(x+1)\port 1)$ defined for $x \in [n,p)$ and undefined otherwise.
Note that the geometry is not expressed by the
names of the cells, with $1$ next to $2$, etc., but by the edges
of the graph. The local rule $f$ is defined 
on disks of radius one as follows (only the significant cases are shown; and the dashed nodes may be present or not):
\begin{center}
\epsfig{file=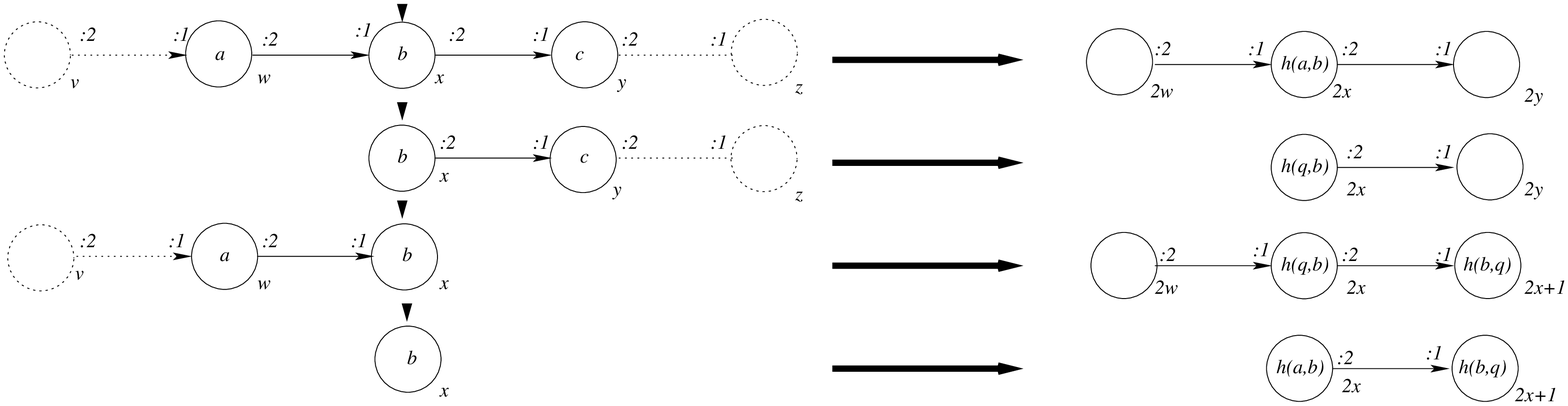,height=3.0cm}
\end{center}
Take $\Sigma=\{0,1\}$, $q=0$ and $h(a,b)=(a+b)\,\mod\,2$. 
Then $c=10011$ is mapped to $c'=110101$. Consider a coding of $c$, 
modulo isomorphism, e.g.
\begin{center}
\epsfig{file=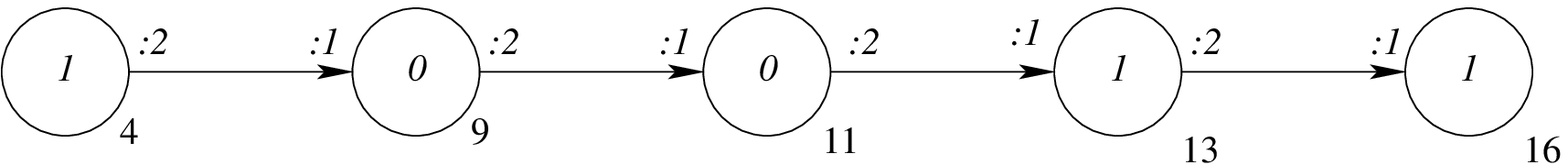,height=0.6cm}
\end{center}
There are five vertices and hence five disks:
\begin{center}
\epsfig{file=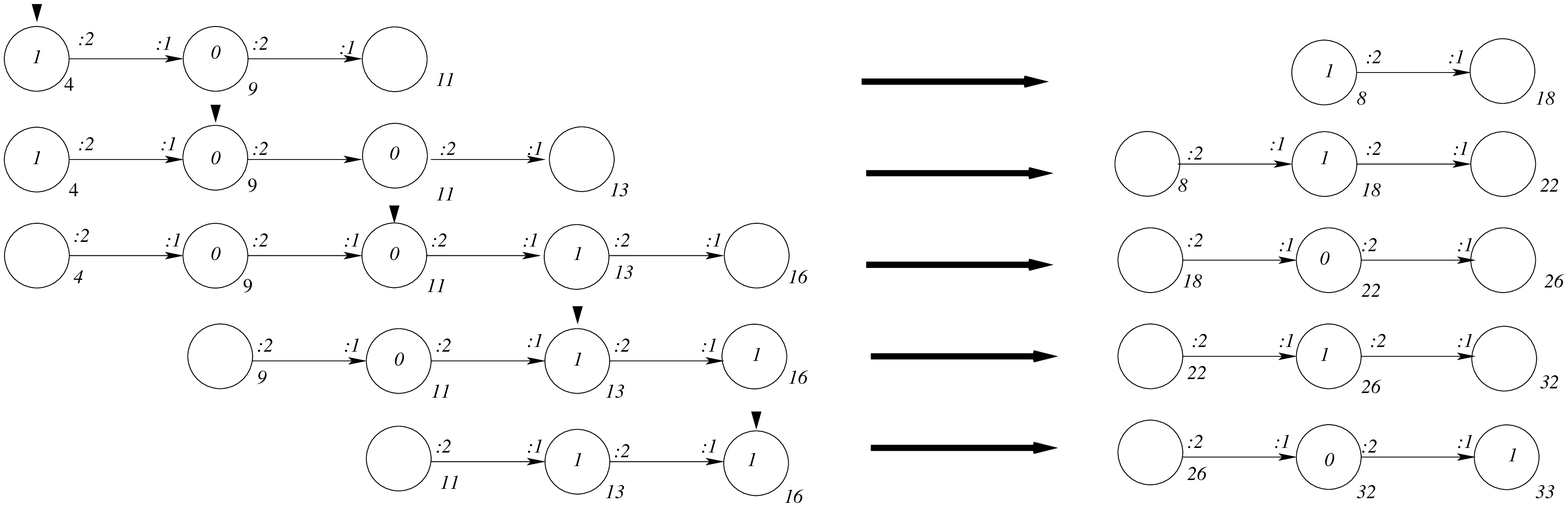,height=3.5cm}
\end{center}
Taking the union, we obtain:
\begin{center}
\epsfig{file=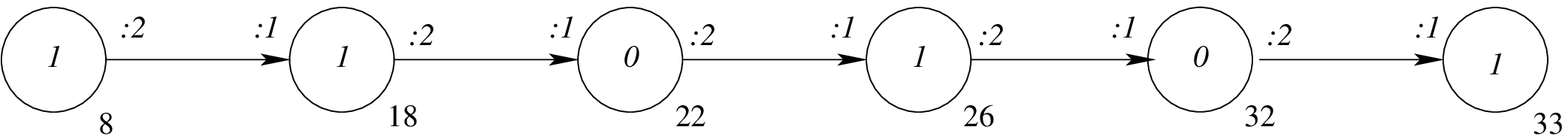,height=0.6cm}
\end{center}
which is indeed a representation of $c'$.

\VL{\begin{proposition}
The function $f$ is a local rule, i.e. a consistent, bounded
dynamics from disks of radius $1$ to graphs.
\end{proposition}}

\VL{\proof{$[\textrm{Dynamics, }(i)]$ Let $R$ be an isomorphism and $R'$ be the function mapping
$2n$ to $2R(n)$ and $2n+1$ to $2R(n) + 1$. It is routine to check that $f \circ R = R' \circ f$. \\
$[\textrm{Dynamics, }(ii)]$ Assume $\bigcap \vt(f(G^{(i)}))\neq \emptyset$ and consider an element $p$ in $\bigcap \vt(f(G^{(i)}))$. If 
$p = 2n$ then $n \in \bigcap \vt(G^{(i)})$ and if $p = 2n + 1$ then $n \in \bigcap \vt(G^{(i)})$. In any case $\bigcap \vt(G^{(i)}) \neq \varnothing$.\\
$[\textrm{Bounded}]$ The graphs of the image of $f$ have at most three vertices.\\
$[\textrm{Consistency}]$
Consider two disks $G_u^1$ and $G_v^1$ centered on
vertices $u$ and $v$.  The domain of the function
$\sigma_1$ giving the state of the vertices of $f(G_u^1)$ is included
in $\{2u, 2u + 1\}$ and that of the function $\sigma_2$ giving the
state of the vertices of $f(G_v^1)$ is included in $\{2v, 2v +
1\}$. If $u \neq v$ then these two sets are disjoint, hence the functions $\sigma_1$
and $\sigma_2$ agree when they are both defined. If $u = v$ then $\sigma_1 =
\sigma_2$ and these two functions agree when they are both defined.\\
Consider two disks $G_u^1$ and $G_v^1$ centered on
vertices $u$ and $v$.  
The functions $\delta_1$ giving the
state of the edges in $f(G_u^1)$ and 
$\delta_2$ giving the
state of the edges in $f(G_v^1)$ agree when they are both defined as they are both partial 
function to the same singleton, i.e. they are either defined or undefined.\\
Finally we have to check that each $a\port i$ appears at most once in 
$dom(\delta_1) \cup dom(\delta_2)$. If $a\port i$ points both to $b\port j$ and $b'\port j'$, then $a=2n$, $i=2$ and $j=j'=1$. In this case:
\begin{itemize}
\item If $b$ and $b'$ are both even then $b = 2q$, $b' = 2q'$, and 
$n\port 2$ was pointing to $q\port 1$ and $q'\port 1$ in $G$. Hence $q = q'$ and so
$b = b'$.
\item If $b$ and $b'$ are both odd then $b = 2q+1$, $b' = 2q'+1$, and
$q = n$, $q' = n$.  Hence $q = q'$ and so $b = b'$.
\item If $b$ is even and $b'$ is odd then $b = 2q$, $b' = 2q'+1$, and 
since $2n\port 2$ points to $2q\port 1$, $n\port 2$ was pointing to $q\port 1$ in 
$G$, moreover since $2n\port 2$ points to $2q+1\port 1$, $n\port 2$ was pointing to no
vertex in $G$. A contradiction.
\end{itemize}
If $a\port i$ is pointed both by $b\port j$ and $b'\port j'$, then $i=1$, $j=j'=2$, $b=2q$ and $b'=2q'$. In this case:
\begin{itemize}
\item If $a$ is odd then $a=2n+1$, $q=q'=n$, and so $b=b'$.
\item If $b$ is even then $a=2n$, and both $q$ and $q'$ were pointing to $n$ in $G$. Hence $q = q'$ and so
$b = b'$.
\end{itemize}
\hfill$\Box$}}

\medskip

\noindent{\em The inflating grid}. In another extreme case,
the graph gets radically modified as each vertex 
$v$ gives rise to four vertices $4v$, $4v+1$, $4v+2$, and $4v+3$. The general
case of the local rule, defined on disks of radius zero, is the following:
\begin{center}
\epsfig{file=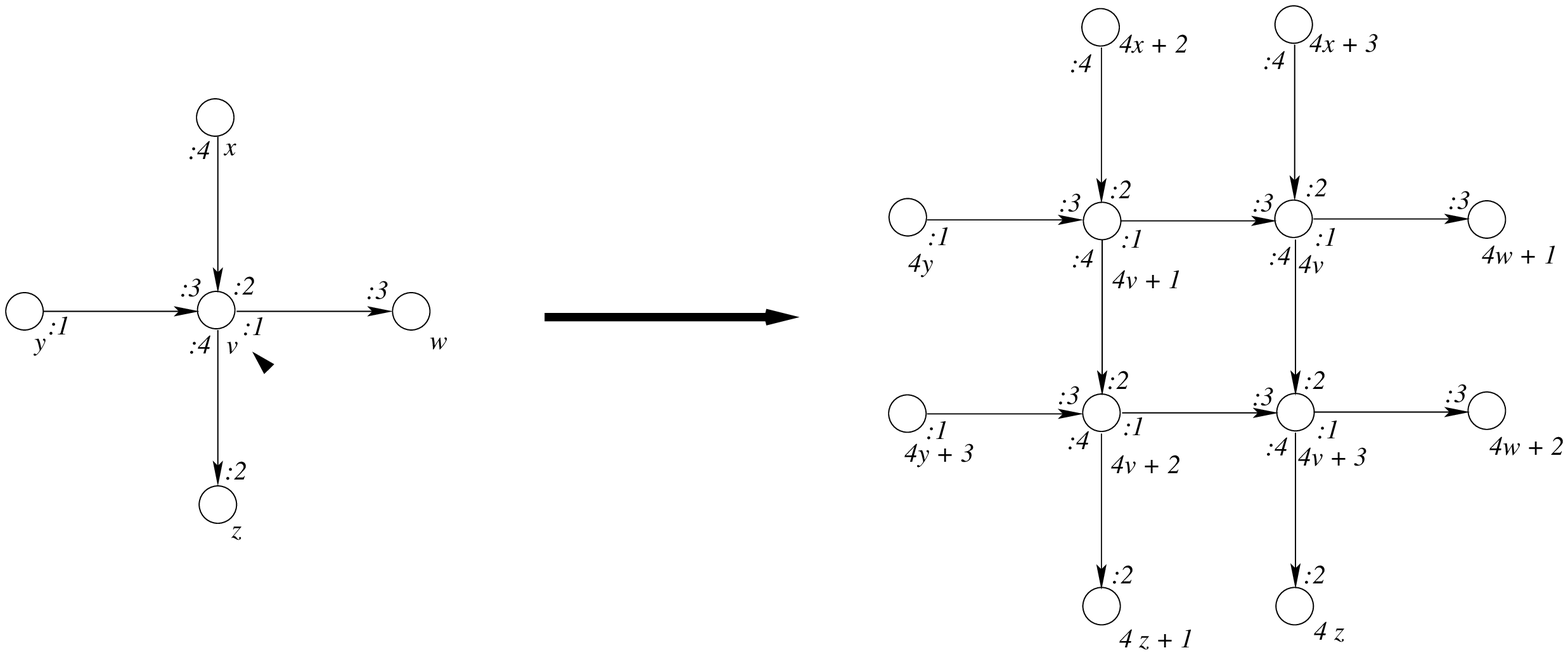,width=8.5cm}
\end{center}
but we have to include fifteen other cases for vertices that do not have 
neighbours in all directions. \VL{For instance :
\begin{center}
\epsfig{file=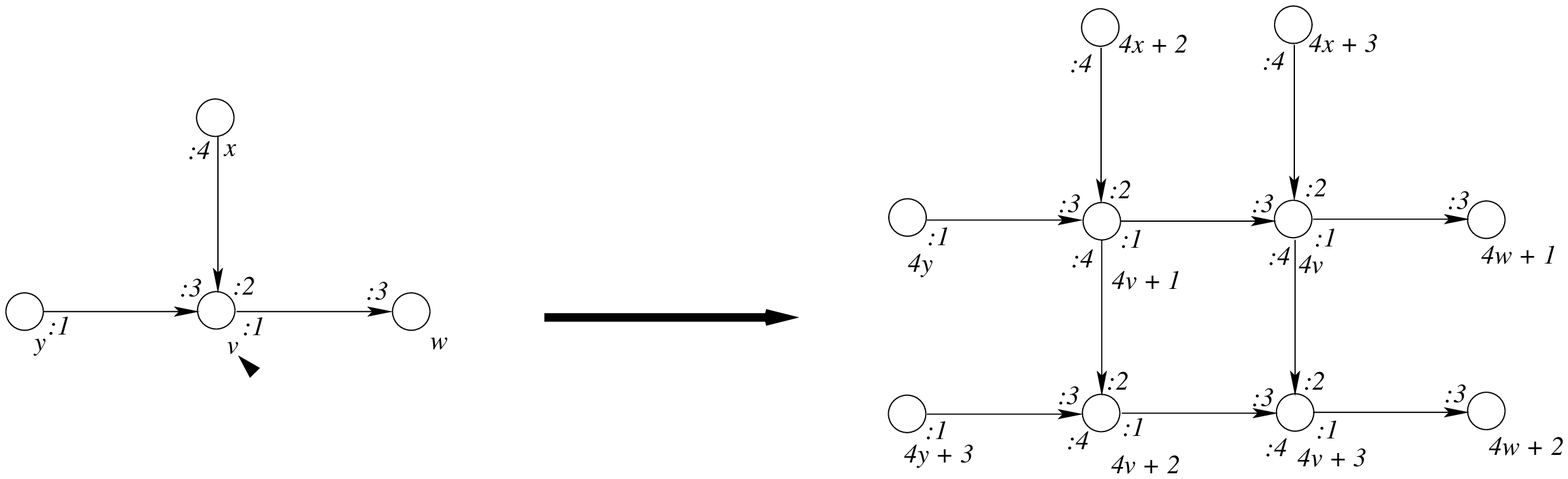,width=8.5cm}
\end{center}}
This inflating grid may also be viewed as a way to generate smaller 
and smaller structures in a fixed size system. 

If we now include states in the guise of colours on vertices, start with a 
grey vertex and rewrite 
\begin{itemize}
\item a black vertex to a cluster of four black vertices,
\item and a grey vertex to a cluster of grey, grey, grey, and black vertices, 
\end{itemize}
we get the picture on the left. 
If, on the other hand, we rewrite 
\begin{itemize}
\item a black vertex to a cluster of four black vertices, 
\item a white vertex to a cluster of white, white, white, and a black vertex,
\item and a grey vertex to a cluster of four white vertices, 
\end{itemize}
we get the picture on the right:
\begin{center}
\hfill
\epsfig{file=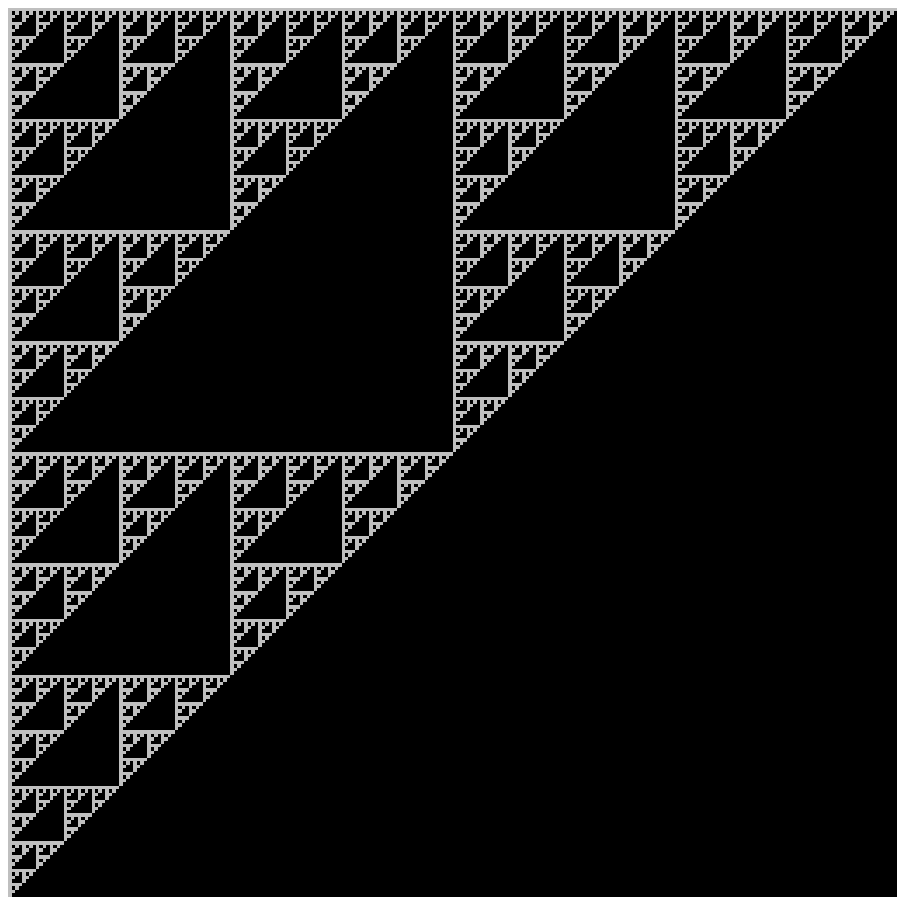,height=4.5cm}
\hfill
\epsfig{file=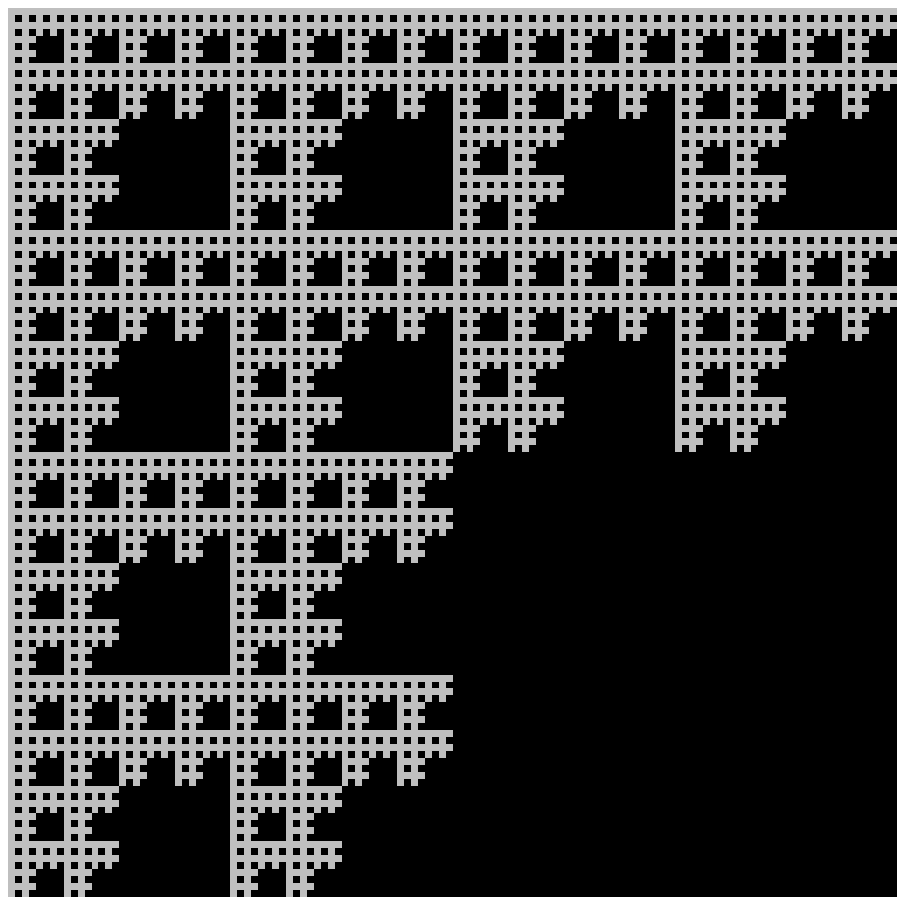,height=4.5cm}
\hfill~
\end{center}
\medskip

\VL{\subsection{Causal is localizable}\label{subsec:causalislocal}}

We now establish our main theorem: that a dynamics is causal
if and only if it is localizable. On the one hand this shows that there is always a
concrete way to construct a causal dynamics. On the other hand, this shows that the notion of a
localizable dynamic is grounded on physical principles such 
as the bounded velocity of the propagation of information. Thus the Physical and the Computer Science-oriented notions coincide. 
A similar theorem is trivial for classical cellular automata, but much more challenging
for instance for reversible cellular automata \cite{KariBlock,Durand-LoseBlock,ArrighiBLOCKREP}, quantum 
cellular automata \cite{ArrighiUCAUSAL}, and graph dynamics. 
In contrast, the extension to probabilistic cellular automata fails
\cite{ArrighiCIE}.

\begin{theorem}[Structure]\label{th:structure}
Let $F$ be a dynamics from ${\cal G}_{\Sigma, \Delta, \pi}$ to ${\cal G}_{\Sigma, \Delta, \pi}$.
$F$ is causal if and only if it is localizable.
\end{theorem}

\VL{\proof{
$[\Leftarrow]$ Suppose $F$ is localizable with local rule $f$. 
We have $F(G)=\bigcup f(G^r_v)$.\\
$[\textrm{Dynamics},\,(i)]$ Consider an isomorphism $R$. Since $f$ is a local rule, it is a dynamics, and so there exists $R'$ such that $f\circ R=R'\circ f$. As a consequence for any $G$ in ${\cal G}_{\Sigma, \Delta, \pi}$:
\begin{align*}
F(RG) &= \bigcup_{v'\in RG} f((RG)^r_{v'}) = \bigcup_{v\in G} f(R(G^r_{v}))\\
&= \bigcup_{v\in G} R'f(G^r_v) = R' \bigcup_{v\in G} f(G^r_v) = R'F(G).
\end{align*}
$[\textrm{Dynamics},\,(ii)]$ Consider some graphs $(G^{(i)})$ having empty intersection. Hence for all vertices $(v_i)$ of these graphs, the corresponding neighbourhoods $(G^{(i)})^r_{v_i}$ have empty intersection. Because $f$ is a dynamics the images of the neighbourhoods $f((G^{(i)})^r_{v_i})$ have empty intersection. As a consequence the images of the graphs $F(G^{(i)})$ have empty intersection.\\
$[\textrm{Causal},\,(i)]$ Consider $a(.)$ the antecedent codynamics of $F$. Consider $v\in a(v'), G, H$ such that $G^{2r}_{v}=H^{2r}_{v}$. Let us call $C^f_G(v')=\{u\;|\;v'\in f(G^r_u)\}$ and similarly for $H$. We have that $F(G)_{v'}=\big(\bigcup_{u\in C^f_G(v')} f(G^r_u)\big)_{v'}$ and similarly for $H$. Using Lemma \ref{lem:antecedentcodynamics} the antecedent $v$ belongs to  every $(G^r_u)_{u\in C^f_G(v')}$ and similarly for $H$. Since their images under the dynamics $f$ have a non-empty intersection, the $G^r_u$'s intersect, and similarly for $H$. As a consequence the fact that $G^{2r}_{v}=H^{2r}_{v}$ guarantees that for all $u$ in $(C^f_G(v')\cup C^f_H(v'))$ we have $G^r_u=H^r_u$ (and hence $f(G^r_u)=f(H^r_u)$). It follows that $C^f_G(v')=C^f_H(v')$ and that $$\bigcup_{u\in C^f_G(v')} f(G^r_u)\;\;=\;\bigcup_{u\in C^f_H(v')} f(H^r_u).$$
Therefore $F(G)_{v'}$ and $F(H)_{v'}$ are equal.\\
$[\textrm{Causal},\,(ii)]$ 
Fix a certain $G$ and $v\in G$. For $v'$ to have antecedent $v$, it must be the case that there exists $u$ such that $v\in G^r_u$, i.e. $u\frown^{\leq r}v$, and $v'\in f(G^r_u)$. In other words, 
$$v'\in\bigcup_{u\in G^r_v} \vt(f(G^r_u)).$$
Due to the boundedness of disks of radius $r$ and of $f$, this is a bounded union of bounded sets.\\
$[\Rightarrow].$ Suppose that $F$ is a causal graph dynamics of radius $r$, and that it has antecedent codynamics $a(.)$. We call $a^{-1}(.)$ the reciprocal function of $a(.)$, i.e. such that $v'\in a^{-1}(v)$ is equivalent to $v\in a(v')$. Consider $v'\in F(G)$. The induced subgraph $F(G)_{v'}$ is a function of $G^r_v$, with $v\in a(v')$. Therefore $F(G)_{a^{-1}(v)}$ is a function of $G^{r}_v$. Hence $F(G)_{a^{-1}(v)}=F(G^{r}_v)_{a^{-1}(v)}$. We call $f$ the function mapping $G^{r}_v$ to $F(G^{r}_v)_{a^{-1}(v)}$. Notice that for all $v'$, $f(G^r_v)_{v'}$ is $F(G)_{v'}$ when $v\in a(v')\cap \vt(G)$, and is empty otherwise.\\
$[\textrm{Dynamics}, (i)]$ We want to prove that for all $R$, $R'\circ f = f\circ R$, with $R'$ a conjugate of $F$ with respect to $F$. Notice that for any $G^{r}_v$:
\begin{eqnarray*}															
R'f(G^{r}_v) &=& R'(F(G)_{a^{-1}(v)})\qquad\textrm{whereas}\\
f(R(G^{r}_v)) &=& f((RG)^{r}_{Rv})=F(RG)_{a^{-1}(Rv)}=(R'F(G))_{a^{-1}(Rv)}=R'(F(G)_{R'^{-1}(a^{-1}(R(v)))}).
\end{eqnarray*}
They are equal since $R'^{-1}\circ a^{-1}\circ R=(a\circ R')^{-1}\circ R=(R\circ a)^{-1}\circ R=a^{-1}\circ R^{-1}\circ R=a^{-1}$, relying on Lemma \ref{lem:antecedentcodynamics}.\\
$[\textrm{Dynamics}, (ii)]$ Notice that by definition, $f(G^{r}_v)\subseteq F(G^{r}_v)$. Thus if $(D^{(i)}_{v_i})$ is a family of disks such that $\bigcap_i \vt(D^{(i)}_{v_i}) = \emptyset$, then since $F$ is a dynamics we have $\bigcap_i \vt(F(D^{(i)}_{v_i})) = \emptyset$, and so $\bigcap_i \vt(f(D^{(i)}_{v_i})) = \emptyset$ as required.\\
$[\textrm{Bounded}]$ The second causality condition states there exists $b$ such that $b\geq|\{v'\in F(G)\,|\,a(v')=v\}|$. Hence for all $v$, $b\geq|vt(F(G)_{a^{-1}(v)})|$. Since $f(G^r_v)$ is defined to be $F(G)_{a^{-1}(v)}$ we have the boundedness of $f$.\\
$[\textrm{Consistent}]$ Consider a graph $G$ and two subdisks $G^{r}_u$, $G^{r}_v$. Then by definition of $f$, both $f(G^{r}_u)$ and $f(G^{r}_v)$ are subgraphs of $F(G)$. Hence they are consistent, and thus $f$ is consistent. Since $f$ is consistent we can define the function $F'$ induced by $f$, i.e. the one mapping $G$ to $\bigcup f(G^{r}_u)$. For all $v'\in F(G)$ we have, by Lemma \ref{lem:antecedentcodynamics}, that 
$\{v\in G\,|\,v\in a(v')\}$ is non-empty by Lemma \ref{lem:antecedentcodynamics}. Hence, 
$$F'(G)_{v'} = \bigcup f(G^r_u)_{v'} = \bigcup_{v\in a(v')\cap \vt(G)} f(G^{r}_v)_{v'} = F(G)_{v'}.$$
Moreover $v'\notin F(G)$ entails that for all $v$, $v'\notin F(G^{r}_v)$, thus $v'\notin f(G^{r}_v)$, and so $v'\notin F'(G)$. Hence $F'(G)=F(G)$. \hfill $\Box$}}

\VC{\section{Properties}}

\VL{\section{Stability}\label{sec:stability}}

\VL{In this section, we prove that causal}\VC{Causal} graph dynamics are closed under
composition.  This is an important indicator of the robustness of this notion.  
Such a result holds trivially for classical and reversible cellular automata, but
depends on the chosen definition for quantum \cite{DurrUnitary,SchumacherWerner,ArrighiLATA} and probabilistic
cellular automata \cite{ArrighiPCA}. \VL{We also show that causal dynamics of radius one are
universal.}

\VL{\begin{lemma}[Past subgraph]\label{lem:pastsubgraph}
Consider $F$ a causal dynamics induced by the local rule $f$ of radius $r$ (i.e. diameter $2r+1$). Consider $G$ in ${\cal G}_{\Sigma, \Delta, \pi}$, $v$ in $G$, $v'$ in $f(G^r_v)$. 
Consider a disk ${F(G)}^{r'}_{v'}$ (i.e. of diameter $d'=2r+1$).
Then this disk is a subgraph of $F(G^{2rr'+r+r'}_v)$. 
The disk $G^{2rr'+r+r'}_v$ has diameter $d''=d'd$.
\end{lemma}}

\VL{\proof{By definition any vertex $u'$ of ${F(G)}^{r'}_{v'}$ is connected to $v'$ by a path $u'=w_1'\frown\ldots\frown w_i'\frown\ldots\frown w_k'=v'$ of length $k-1$ less or equal to $r'$. In the worst case this path is of length $r'=k-1$, and each vertex $w_i'$ belongs to a distinct $f(G^r_{v_i})$, with $v_0=u$ and $v_k=v$. The question is whether $G^r_{u}$ is a subgraph of $G^{2rr'+r+r'}_v$. To this end, notice that any two vertices of $G^r_{v_i}$ are connected by a path of length at most $2r$ and that any vertex of $G^r_{v}$ is connected to $v$ by a path of length at most $r$. Moreover according to Theorem \ref{th:structure}, the fact that $w_i'$ is connected to $w_{i+1}'$ implies that $G^r_{v_i}$ is connected with $G^r_{v_{i+1}}$. As a consequence, the vertex $u$ is, in the worst case, connected with $v$ by a path of length
$2rr'+r+r'$, where the first subterm of the sum is the path length across the first $r'$ subgraphs $G^r_{v_i}$, the second subterm is the path length inside the last subgraph, and the third subterm is the path length between the subgraphs .\hfill$\Box$}}

\begin{theorem}[Composability]\label{th:composability}
Consider $F_1$ a causal dynamics induced by the local rule $f_1$ of radius $r_1$ (i.e. diameter $d_1=2r_1+1$). Consider $F_2$ a causal dynamics induced by the local rule $f_2$ of radius $r_2$ (i.e. diameter $d_2=2r_2+1$). Then $F_2\circ F_1$ is a causal dynamics induced by the local rule $g$ of radius $r''=2r_1r_2+r_1+r_2$ (i.e. diameter $d''=d_1d_2$) from ${\cal D}^{r''}$ to ${\cal G}_{\Sigma, \Delta, \pi}$ which maps $N^{r''}_v$ to
$$\bigcup_{v'\in f_1(N^{r_1}_v)} f_2\Big(\big(\bigcup_{v \in N^{r''}_v} f_1(N^{r_1}_v)\big)^{r_2}_{v'}\Big)$$
\end{theorem}

\VL{\proof{
\begin{eqnarray*}
\bigcup_{v\in G} g(G^{r''}_v)&=&
\bigcup_{v\in G,\,v'\in f_1((G^{r''}_v)^{r_1}_v)}
f_2\Big(\big(\bigcup_{v \in G^{r''}_v} f_1((G^{r''}_v)^{r_1}_v)\,\,\,\big)^{r_2}_{v'}\Big)\\
&=&
\bigcup_{v\in G,\,v'\in f_1(G^{r_1}_v)} 
f_2\Big(\big(\bigcup_{v \in G^{r''}_v} f_1(G^{r_1}_v)\,\,\,\big)^{r_2}_{v'}\Big)\\
&=&
\bigcup_{v\in G,\,v'\in f_1(G^{r_1}_v)} 
f_2\Big(\big(F_1(G)\big)^{r_2}_{v'}\Big)\quad\textrm{by Lemma \ref{lem:pastsubgraph}}\\
&=&
\bigcup_{v'\in F_1(G)} 
f_2\Big(\big(F_1(G)\big)^{r_2}_{v'}\Big)\quad =\quad F_2\big(F_1(G)\big)
\end{eqnarray*}
\hfill$\Box$}}

\VC{We also show that causal dynamics of radius one are
universal.}
\begin{proposition}[Universality of radius one]\label{prop:closure}
Consider $F$ a causal dynamics of radius $r=2^l$ over ${\cal G}_{\Sigma, \Delta, \pi}$. There exists $F'$ a causal dynamics of radius $1$ over ${\cal G}_{\Sigma^{l+1},\Delta\cup\{*\},\pi^r}$, such that ${F'}^{l+1}=F$.
\end{proposition}

\VL{\proof{Outline. 
Over the first $i=1\ldots l$ steps, the neighbours of radius $2^l$ are computed. More precisely, states of vertices are left identical, whereas an ancillary edge with state $*$ is added between any two previously connected vertices. Moreover, the vertices count until stage $l$. At this point, the neighbours that were initially of radius $r$ have become visible within a radius one. The local rule of $F$ can apply, all ancillary edges be dropped, and counters be reset.\hfill$\Box$}}

\VL{\section{Causal and continuous dynamics}\label{sec:continuity}}

The notion of causality (see Definition \ref{def:causal}), 
is based on the mathematical notion of uniform continuity: the radius $r$ is
independent of the vertex $v$ and of the graph $G$. It is well-known that
in general uniform continuity implies continuity, and that on compact spaces
continuity implies uniform continuity. Such results have been
extended to cellular automata \cite{Hedlund}. \VL{We show here that they also extend to}\VC{This is also true of} graph dynamics. Uniform continuity always implies continuity, and the 
converse holds when the state spaces are finite.

\VL{\subsection{General case}}

We now develop a notion of continuity, and find out that it is equivalent to limit-preservation. 

\begin{definition}[Continuous dynamics]
A dynamics $F$ from ${\cal G}_{\Sigma, \Delta, \pi}$ to ${\cal G}_{\Sigma, \Delta, \pi}$ is {\em continuous} if and only if:
\begin{eqnarray*}
\forall r',\forall v', v\in a(v'), \forall G, \exists r, \forall H,\quad \big[G^r_{v}=H^r_{v}\Rightarrow F(G)^{r'}_{v'}=F(H)^{r'}_{v'}\,\big]
\end{eqnarray*}
with $a(.)$ the antecedent codynamics of $F$.
\end{definition}

\begin{definition}[Limit at $A$]
Consider a function $r\mapsto (G(r),A)$ from $\N$ to ${\cal P}_{\Sigma, \Delta, \pi}$. We say that it converges to $(G,A)$ if and only if for all $r$ there exists $s$ such that for all $s'\geq s$, $G(s')^r_{A}=G^r_A$. 
\end{definition}

\begin{proposition}[Continuity as limit preservation]
Consider a dynamics $F$ having antecedent codynamics $a(.)$. $F$ is continuous if and only if it preserves limits, meaning that if the function $s\mapsto (G(s),A)$ converges to $(G,A)$, then the function $s\mapsto (F(G(s)),a^{-1}(A))$ converges to $(F(G),a^{-1}(A))$.
\end{proposition}

\VL{\proof{
$[\Rightarrow]$ Suppose $F$ continuous and having antecedent codynamics $a(.)$. Suppose $s\mapsto (G(s),A)$ converges to $(G,A)$. 
Let $k$ be an integer. Continuity tells us that there exists $r_k$ such that for all $H$, 
$$\big[G^{r_k}_{A}=H^{r_k}_{A}\Rightarrow F(G)^{k}_{a^{-1}(A)}=F(H)^{k}_{a^{-1}(A)}\,\big].$$
Hence there exists $s$ such that for all $s'\geq s$, $G^{r_k}_{A}=G(s')^{r_k}_{A}$, and taking $G(s')$ for $H$, we obtain $F(G)^{k}_{a^{-1}(A)}=F(G(s'))^{k}_{a^{-1}(A)}$ as needed.\\
$[\Leftarrow]$ By contradiction, suppose that $F$ preserves limits, but $F$ fails to be continuous. The fact that $F$ is not continuous entails the existence of $A$, $r'$, $G$, such that for all $r$, there exists $H(r)$ such that $G^r_{A}=H(r)^r_{A}$, and yet $F(G)^{r'}_{a^{-1}(A)}\neq F(H(r))^{r'}_{a^{-1}(A)}$. Fix such $A$, $r'$, $G$, and $H(.)$. Since $H(r)^r_{A}=G^r_{A}$, by limit preservation we should have for all $k$, there exists $s_k$ such that, $F(H(s_k))^k_{a^{-1}(A)}=F(G)^k_{a^{-1}(A)}$. Taking $r'$ for $k$ yields $F(H(s_{r'}))^{r'}_{a^{-1}(A)}=F(G)^{r'}_{a^{-1}(A)}$, a contradiction. \hfill$\Box$}}

\begin{corollary}[Causality implies limit preservation]\label{cor:}
If a dynamics $F$ is causal then it is uniformly continuous, hence it is continuous and it preserves limits.
\end{corollary}
\VL{\proof{Continuity is clearly a weakened form of Condition $(i)$ of causality. \hfill $\Box$}}

\noindent The converse is not true in general, and will now be investigated in the finite case.
\VL{\subsection{Finite case}\label{subsec:finiteproperties}}
\VC{For the next two theorems}\VL{In this entire section} the graphs may still be infinite in size, but their set of states of vertices and edges are supposed to be finite. Both suppositions are necessary to obtain the compactness property, which works modulo isomorphism.

\VL{\begin{lemma}[Compactness]\label{lem:compactness}
Consider the set of pointed graphs ${\cal P}_{\Sigma, \Delta, \pi}$ with $\Sigma$ and $\Delta$ finite. Then for any sequence of pointed graphs $(r\mapsto (G(r),v(r)))$, there exists a pointed graph $(G,v)$, an increasing subsequence $(k\mapsto r_k)$ with $r_k\geq k$, and a sequence of isomorphisms $(k\mapsto R(k))$, such that for all $k$ we have $(R(k)G(r_k))^k_v=G^k_v$, with $R(k)(v(r_k))=v$.
\end{lemma}}

\VL{\proof{Outline.
Because $\Sigma$ and $\Delta$ are finite, and there is an infinity of $(G(r),v(r))$, there must exist a pointed graph of radius zero $G^0_v$ such that there is an infinity of $(G(r),v(r))$ having $G(r)^0_{v(r)}\approx G^0_v$. Choose one of them to be $(G(r_0),v(r_0))$. Let $R(0)$ be such that $R(0)(G(r_0)^0_{v(r_0)})=G^0_v$. Fix the neighbourhood of radius zero around $v$ of $G$ to be precisely this $G^0_v$. Now iterate: because the degree of the graph is bounded by $\pi$, and because $\Sigma$ and $\Delta$ are finite but there is an infinity of $(G(r))$ having the above property, there must exist a pointed graph of radius one $G^1_0$ such that $(G^1_v)^0_v=G^0_v$ and such that there is an infinity of $(G(r),v(r))$ having $G(r)^1_{v(r)}\approx G^1_0$. Choose one of them as $(G(r_1),v(r_1))$. Let $R(1)$ be such that $R(1)(G(r_1)^1_{v(r_1)})=G^1_v$. Fix the neighbourhood of radius one around $v$ of $G$ to be precisely this $G^1_v$. Etc. 
\hfill$\Box$}}

\noindent In topology, continuity and compactness entail uniform continuity. We do not have a clear topology for our graphs but a similar result holds.
\begin{theorem}[Continuity and causality]\label{th:continuity}
Consider $F$ a dynamics from ${\cal G}_{\Sigma, \Delta, \pi}$ to ${\cal G}_{\Sigma, \Delta, \pi}$, with $\Sigma$ and $\Delta$ finite. If $F$ is continuous, then it verifies the first causality condition, and conversely. 
\end{theorem}

\VL{\proof{$[\Leftarrow]$ See Corollary \ref{cor:}.\\
$[\Rightarrow]$ 
By contradiction: suppose that $F$ is continuous dynamics, but that it is not causal. Say it has antecedent codynamics $a(.)$.\\
$[\textrm{Test graphs}].$ 
Non-causality gives: $\forall r, \exists v(r), \exists G(r), H(r),$
$$\big[G(r)^{r}_{v(r)}=H(r)^{r}_{v(r)}\wedge F(G(r))_{a^{-1}(v(r))}\neq F(H(r))_{a^{-1}(v(r))}\big].$$
Fix some sequences $r\mapsto v(r)$, $r\mapsto G(r)$ and $r\mapsto H(r)$ verifying the above.\\
$[\textrm{Limit of test graphs}].$ Using the finiteness of $\Sigma$ and $\Delta$ and the construction of Lemma \ref{lem:compactness}, there exists a pointed graph $(G,v)$, an extracted subsequence $(k\mapsto r_k)$, and a sequence of isomorphisms $(k\mapsto R(k))$ such that for all $k$ we have $(R(k)G(r_k))^k_v=G^k_v$, with $R(k)(v(r_k))=v$. Fix such $G$ and the sequences $(k\mapsto r_k)$ and $(k\mapsto R(k))$.\\
$[\textrm{Continuity at the limit}].$ We have $(R(k))^{-1} G^k_v=G(r_{k})^k_{v(r_k)}=H(r_{k})^k_{v(r_k)}$ and thus 
$(R(k)G(r_{k}))^k_v=G^k_v=(R(k)H(r_{k}))^k_v$. Applying continuity at $v$ and $G$ yields that there exists $k$, such that we have both 
\begin{align*}
&[G^{k}_{v}=(R(k)G(r_k))^{k}_{v}\;\Rightarrow\; F(G)_{a^{-1}(v)}=F(R(k)G(r_k))_{a^{-1}(v)}\big]\\
\textrm{ and }&[G^{k}_{v}=(R(k)H(r_k))^{k}_{v}\;\Rightarrow\; F(G)_{a^{-1}(v)}=F(R(k)H(r_k))_{a^{-1}(v)}\big].
\end{align*}
Fix such $k$. This yields $F(R(k)G(r_k))_{a^{-1}(v)}=F(R(k)H(r_k))_{a^{-1}(v)}$.
However,
$$F(R(k)G(r_k))_{a^{-1}(v)}=R'(k) (F(G(r_k))_{R'(k)^{-1}(a^{-1}(v))}) =R'(k) (F(G(r_k))_{a^{-1}(v(r_k))}),$$
where we used $(R'(k)^{-1}\circ a^{-1})=(a\circ R'(k))^{-1}=(R(k)\circ a)^{-1} = (a^{-1}\circ R(k)^{-1})$ as given by Lemma \ref{lem:antecedentcodynamics}.
Similarly, $F(R(k)H(r_k))_{a^{-1}(v)}=R'(k) (F(H(r_k))_{a^{-1}(v(r_k))})$. Thus $R'(k) (F(G(r_k))_{a^{-1}(v(r_k))})= R'(k) (F(H(r_k))_{a^{-1}(v(r_k))})$. This entails that $F(G(r_k))_{a^{-1}(v(r_k))}=F(H(r_k))_{a^{-1}(v(r_k))}$. A contradiction.\hfill$\Box$}}

\VL{\section{Invertibility}\label{sec:invertibility}}

A causal dynamics is said to be {\em invertible} if it has an inverse.
It is said to be {\em reversible} if this inverse is itself a causal
dynamics. \VL{We prove that, as}\VC{As} for the cellular automata, invertibility
implies reversibility: if a causal dynamics has an inverse, then this
inverse is also a causal dynamics. \VL{As in the previous section, the
results only holds when the state space is finite.} The proof of this
invertibility theorem is similar to the proof that continuity 
implies uniform continuity, both proceeding by extracting converging 
subsequences from arbitrary ones.

\VL{\subsection{General case}}

\VL{Without even supposing causality there is a few things we can say about 
invertible dynamics.}
\begin{definition}[Invertible dynamics]
A dynamics $F$ from ${\cal G}_{\Sigma, \Delta, \pi}$ to ${\cal G}_{\Sigma, \Delta, \pi}$ is
{\em invertible} if and only if there exists a dynamics $F^\dagger$ such that $F^\dagger \circ  F=Id$ and $F\circ F^\dagger=Id$. This inverse is unique.
\end{definition}

\VL{\noindent Indeed, invertible dynamics cannot take a graph apart.}

\VL{\begin{lemma}[Invertible implies connected-preserving]\label{lem:connectedpreserving}
If a dynamics $F$ is invertible, then it is connected-preserving, meaning that for any two graphs $G$ and $H$, 
$$\big[\vt(G)\cap \vt(H) =\emptyset\;\Leftrightarrow\;\vt(F(G))\cap \vt(F(H))=\emptyset\;\big].$$
\end{lemma}}

\VL{\proof{$[\Rightarrow]$ This is the dynamicity of $F$.\\
$[\Leftarrow]$ By contradiction, suppose two graphs $G$ and $H$ such that $\vt(G)\cap \vt(H) \neq\emptyset$, but $\vt(F(G))\cap \vt(F(H))=\emptyset$. Take $v\in (\vt(G)\cap \vt(H))$. Therefore $v\in F^\dagger(F(G))$ and $v\in F^\dagger(F(H))$, and so $\vt(F^\dagger(F(G)))\cap \vt(F^\dagger(F(H)))\neq\emptyset$, which is contradictory with the dynamicity of $F^\dagger\circ F$. \hfill$\Box$}}

\VL{
\noindent Moreover, invertible dynamics must preserve vertices up to an isomorphism.
\begin{lemma}[Invertible implies vertex-preserving up to]\label{lem:vertexpreservingupto}
If a dynamics $F$ is invertible with inverse $F^\dagger$, then:
\begin{align*} 
\forall G,\quad |\vt(G)|&=|\vt(F(G))|\\
\forall v',\quad |a(v')|&=1\\
a^{\dagger}\circ a = a\circ a^{\dagger} &= Id
\end{align*} 
with $a$ (resp. $a^\dagger$) the antecedent codynamics of $F$ (resp. $F^\dagger$).
\end{lemma}}

\VL{\proof{First we show that if $v'\in a^\dagger(v)$ then $a(v')=\{v\}$. 
Assume $v'\in a^\dagger(v)$ and take $w\in a(v')$. We prove $w=v$. Indeed, we have
$\forall H,\,[v\in F^\dagger(H) \Rightarrow v'\in H]$. Apply this to $H=F(\{v\})$. This gives $v'\in F(\{v\})$.  Moreover, we have 
$\forall G,\,[v'\in F(G) \Rightarrow w\in G]$. Apply this to $G=\{v\}$. This gives $[v'\in F(\{v\}) \Rightarrow w=v]$. Combining both, $w=v$. Thus $a(v')\subseteq \{v\}$ and as it is non-empty $a(v')=\{v\}$.\\
Second we show that if $v\in a(v')$ then $a^\dagger(v)=\{v'\}$. This holds by symmetry.\\
Finally, take any $v'$ and $v\in a(v')$. By the second fact, $v'\in a^\dagger(v)$. Hence by the first fact, $a(v')=\{v\}$. 
\hfill$\Box$}}

\VL{
\noindent In fact, this isomorphism can be undone.
\begin{lemma}[Invertible implies vertex-preserving]\label{lem:vertexpreservingstrict}
Consider $F$ an invertible dynamics with antecedent codynamics $a$. 
Let $\ovb{F}$ be equal to $(a\circ F)$. We have that $\ovb{F}$ is an invertible dynamics such that:
\begin{align*} 
\forall R,\quad R\circ \ovb{F} &= \ovb{F}\circ R \\
\forall G,\quad \vt(G)&=\vt(F(G))\\
\ovb{a} = \ovb{a}^{\dagger} &= Id
\end{align*} 
with $\ovb{a}$ (resp. $\ovb{a}^\dagger$) the antecedent codynamics of $\ovb{F}$ (resp. $\ovb{F}^\dagger$).
\end{lemma}}

\VL{\proof{Since $F$ is invertible, we have from Lemma \ref{lem:vertexpreservingupto} that $a$ is an isomorphism. Hence $a$ is an invertible dynamics. Hence $\ovb{F}$ is an invertible dynamics as the composition of two invertible dynamics. It is commuting because, for all $R$, $\ovb{F}\circ R = a\circ F\circ R = a\circ R'\circ F = R\circ a \circ F = R\circ \ovb{F}$, where we used Lemma \ref{lem:antecedentcodynamics}. Since $\ovb{F}$ is commuting and invertible, it does not allow the introduction of new vertices by Lemma \ref{lem:commutingimpliesnonincreasing}, and it is vertex-preserving by Lemma \ref{lem:vertexpreservingupto}, therefore it leaves vertices unchanged, i.e. $\vt(G)=\vt(\ovb{F}(G))$. Now take $v\in \ovb{a}(v')$. This entails that for all $G$, $v'\in \ovb{F}(G)$ implies $v\in G$. In particular for $G=\{v'\}$ we get that $v'\in \ovb{F}(\{v'\})$ implies $v=v'$. But since $\vt(\ovb{F}(\{v'\}))=\{v'\}$, we have $v=v'$. So $\ovb{a}$ is indeed the identity. \hfill$\Box$}}

\VL{\subsection{Finite case}}

\VL{Let us now consider causal, invertible dynamics, and show that their inverse is also causal. This property is often referred to as `reversibility'.}
\begin{theorem}[Causal, invertible, reversible]\label{th:reversibility}
Consider $F$ a causal dynamics from ${\cal G}_{\Sigma, \Delta, \pi}$ to ${\cal G}_{\Sigma, \Delta, \pi}$, with $\Sigma$ and $\Delta$ finite. If $F$ is invertible with dynamics $F^\dagger$, then $F^\dagger$ is also causal.
\end{theorem}

\VL{\proof{$[\textrm{Reduction to}\ \ovb{F}]$ Say $F$ has antecedent codynamics $a$, which is an isomorphism by Lemma \ref{lem:vertexpreservingupto}. Then by Lemma \ref{lem:vertexpreservingstrict}, $\ovb{F}=(a\circ F)$ is an invertible dynamics. It is also causal, as the composition of two causal dynamics, by Proposition \ref{th:composability}. Next we will prove the theorem for $\ovb{F}$, i.e. we will prove that $\ovb{F}^\dagger$ is causal. From there it will follow that $F^\dagger= F^\dagger\circ a^\dagger\circ a = (a\circ F)^\dagger \circ a = \ovb{F}\circ a$ is a causal dynamics as the composition of two causal dynamics, by Proposition \ref{th:composability}. Hence we will have proven the theorem in the general case.

\noindent $[\textrm{Uniform continuity}]$ By contradiction, suppose causality of $\ovb{F}$, existence of $\ovb{F}^{\dagger}$, but non-uniform-continuity of this $\ovb{F}^{\dagger}$.

\noindent $[\textrm{Test graphs}].$ 
Non-uniform-continuity of $\ovb{F}^\dagger$ is equivalent to non-continuity of $\ovb{F}^\dagger$ by Theorem \ref{th:continuity}, hence: 
$\forall r, \exists v, \exists G', \exists H'(r),$
$$\big[G'^{r}_{v}=H'(r)^{r}_{v}\;\wedge\; \ovb{F}^\dagger(G')_{v}\neq \ovb{F}^\dagger(H'(r))_{v}\big].$$
where we used $\ovb{a} = \ovb{a}^{\dagger} = Id$ from Lemma \ref{lem:vertexpreservingstrict}. Fix such $v$, $G'$ and $(r\mapsto H'(r))$.

\noindent $[\textrm{Limit of test graphs}].$ 
Let us call $G=\ovb{F}^\dagger(G')$ and $H(r)=\ovb{F}^\dagger(H'(r))$. Using the finiteness of $\Sigma$, $\Delta$, and the construction of Lemma \ref{lem:compactness}, there exists a pointed graph $(H,v)$, an extracted subsequence $(k\mapsto r_k)$, and a sequence of isomorphisms $(k\mapsto R(k))$, such that for all $k$ we have both $(R(k)H(r_k))^k_v=H^k_v$ with $R(k)(v)=v$. Fix such $H$, and the sequences $(k\mapsto r_k)$ and $(k\mapsto R(k))$.\\

\noindent $[\textrm{Uniform continuity at the limit}].$ Since $\ovb{F}$ is causal there exists $(l\mapsto r_{k_l})$ such that $\ovb{F}(R(k)H(r_{k_l}))^l_{v}=\ovb{F}(H)^l_{v}$, where we used $\ovb{a} = \ovb{a}^{\dagger} = Id$ from Lemma \ref{lem:vertexpreservingstrict}. 
This entails that $\ovb{F}(H(r_{k_l}))^l_{v}=\ovb{F}(R(k)^{-1}H)^l_{v}$, where we used $\forall R,\ R\ovb{F} = \ovb{F}R$ from Lemma \ref{lem:vertexpreservingstrict}.

\noindent $[$Relation between $G$ and $S(l)H].$ We have:
\begin{eqnarray*}
G&=&\ovb{F}^\dagger(G')\\
\ovb{F}^\dagger(G')_v&\neq & \ovb{F}^\dagger(H'(r_{k_l}))_v\\
\ovb{F}^\dagger(H'(r_{k_l}))&=&H(r_{k_l})\\
H(r_{k_l})^{k_l}_v&=&R(k_l)^{-1}H^{k_l}_v
\end{eqnarray*}
Let $S(l)=R(k_l)^{-1}$. Since $R(k_l)(v)=v$, we have $S(l)(v)=v$.
Combining this with the above equations we obtain: $G_v\neq S(l)H_v$.

\noindent $[$Relation between $\ovb{F}(G)$ and $\ovb{F}(S(l)H)].$ We have:
\begin{eqnarray*}
\ovb{F}(G) &=& \ovb{F}(\ovb{F}^\dagger(G'))\\
\ovb{F}(\ovb{F}^\dagger(G')) &= & G'\\
G'^{r_{k_l}}_{v} &=& H'(r_{k_l})^{r_{k_l}}_{v}\\
H'(r_{k_l}) &=& \ovb{F}(\ovb{F}^\dagger(H'(r_{k_l}))\\
\ovb{F}(\ovb{F}^\dagger(H'(r_{k_l}))) &=& \ovb{F}(H(r_{k_l}))\\
\ovb{F}(H(r_{k_l}))^l_{v} &=& \ovb{F}(R(k_l)^{-1}H)^l_{v}\\
\end{eqnarray*}
Moreover from Lemma \ref{lem:vertexpreservingstrict}, $\ovb{F}\circ S(l)=S(l)\circ \ovb{F}$.
Combining this with the above equations we obtain: $\ovb{F}(G)^l_{v}=S(l)\ovb{F}(H)^l_{v}$. 

\noindent $[\textrm{Convergence of }S(l)\textrm{ to }T].$ 
Notice that since $S(l+1)\ovb{F}(H)^{l+1}_{v}=\ovb{F}(G)^{l+1}_{v}$, it follows that $S(l+1)\ovb{F}(H)^{l}_{v}=\ovb{F}(G)^{l}_{v}$. 
But since $S(l)\ovb{F}(H)^{l}_{v}=\ovb{F}(G)^{l}_{v}$ it follows that $S(l+1)$ and $S(l)$ coincide on $\vt(\ovb{F}(H)^{l}_{v})$.
We now define the function $T$ from $\vt(\ovb{F}(H))$ to $\vt(\ovb{F}(G))$ such that for all $u\in \ovb{F}(H)$, we choose any $l$ such that $u\in \ovb{F}(H)^l_v$ and let $T(u)=S(l)(u)$.
The function $T$ is injective since $S(l)$ is injective, and so taking distinct $u,w\in \ovb{F}(H)^l_v$ implies that $S(l)(u)$ and $S(l)(w)$ are distinct.
The function $T$ is surjective since for every $u'$ in $\ovb{F}(G)^l_v$ there exists $u$ such that $S(l)(u)=u'$.
So $T$ is bijective from $\vt(\ovb{F}(H))$ to $\vt(\ovb{F}(G)$. Moreover $|N\setminus\{\vt(\ovb{F}(H))\}|$ and $|N\setminus\{\vt(\ovb{F}(H))\}|$ have the same cardinal. 
Hence the bijection $T$ can be extended to become a bijection over the entire set of names $V$, which we now assume.

\noindent $[\textrm{Properties of }T].$ 
Since $\vt(H_v)\subseteq \vt(H) = \vt(F(H))$, there exists $m$ such that $\vt(H_v)\subseteq \vt(F(H)^m_v)$.
Moreover, since $T$ coincides with $S(m)$ over $\ovb{F}(H)^m_v$, it coincides over $\vt(H_v)$, and from $G_v\neq S(m)H_v$
it follows that $G_v\neq T H_v$. Moreover, since $T$ coincides with $S(l)$ over $\ovb{F}(H)^l_v$, we have that for all $l$, $\ovb{F}(G)^l_{v}=T\ovb{F}(H)^l_{v}$.

\noindent $[\textrm{Non-injectivity of }\ovb{F}].$ From $G_v\neq T H_v$ we have $G\neq T H$. From the fact that for all $l$, $\ovb{F}(G)^l_{v}=T\ovb{F}(H)^l_{v}$ and the fact that $\ovb{F}$ is connected-preserving by Lemma \ref{lem:connectedpreserving}, we have $\ovb{F}(G)=T \ovb{F}(H)$, and hence $\ovb{F}(G)=\ovb{F}(T H)$ by Lemma \ref{lem:vertexpreservingstrict}. This contradicts the non-injectivity of $\ovb{F}$.

\noindent $[\textrm{Boundedness}]$ This is a consequence of Lemma \ref{lem:vertexpreservingstrict}.
\hfill $\Box$}}

\VL{\section{Summary and future work}\label{sec:conclusion}}

\VC{\section{Future work}\label{sec:conclusion}}

\VL{\noindent {\em Summary.} The purpose of this paper was to study causal graph dynamics. We did so in a discrete-space, discrete-time, deterministic setting. Let us summarize some of the main ideas and results, but in a different order. By causal, we mean that information cannot propagate faster than some fixed speed bound $r$. We formalized this concept in an axiomatic manner, by stating that the internal state and connectivity $F(G)_{v'}$ of a vertex $v'$ of $F(G)$ must be a function of $G^r_v$ the disk of radius $r$ around some antecedent $v\in a(v')$ of $G$. It is important that an axiomatization be general enough to encompass every possible case, but at the same time concrete enough not to remain a purely abstract mathematical object. Here, the definition was shown to be equivalent to the more operational form $F(G)=\bigcup_v f(G^r_v)$, where $f$ is a local rule that gets applied in parallel everywhere, generating the subgraphs $f(G^r_v)$, then glued together by the union. Notice that not all graphs can be glued together, some consistency conditions are required: we characterized the local rules $f$ that induce causal dynamics $F$. Notice also that in order to glue these graphs, we need to identify some vertices of $f(G^r_u)$ with those of a nearby $f(G^r_v)$, perhaps avoiding collisions with the vertices of some far-away $f(G^r_w)$, say. Hence, vertices need be named, and the way that $f$ generates new names from former names was made to respect the somewhat natural condition that $\bigcap \vt(G^{(i)})=\emptyset$ implies $\bigcap \vt(f(G^{(i)}))=\emptyset$. Moreover, the fact that vertices are named ought not have an excessive impact over $F$, whose behaviour ought to be roughly the same as names get changed through some injective function $R$. This led us to impose the second condition that there always be some $R'$ such that $R'\circ F=F\circ R$. In fact, the graph dynamics considered in this paper were restricted to respect these two conditions outright. This good behaviour of graph dynamics turns out to provide a robust notion of antecedent codynamics $a(.)$: the antecedents of the vertex name $v'$ are those vertex names $v$ which are systematically present in the graphs $G$ that generate this name.\\
The notion of causal graph dynamics was then instantiated into two concrete examples, which can be viewed as limiting cases: Cellular automata feature complex internal state manipulations, whereas the Inflating grid is an extreme case of time-varying neighbourhood. The robustness of the proposed notion was also ascertained by some mathematical, stability results: the composition of two causal graph dynamics is a causal graph dynamics whose local rule we can describe, moreover the speed bound $r=1$ is universal. The credit of Cellular automata theory owes to such foundational results, as well as others, such as the Curtis-Hedlund-Lyndon theorem, which characterizes them a translation-invariant, continuous functions for the Cantor topology on words. This theorem works only for the space of infinite configurations, with cells taking their states in a finite alphabet. We similarly considered infinite graphs with vertices and edges having their states in a finite alphabet. No clear topology was at our disposal, but a notion of limit, which served to define continuity equally well, and obtain an equivalent of the theorem. From similar, compactness arguments, we also obtained that if a causal graph dynamics has an inverse dynamics, then this inverse dynamics is again causal.

\noindent {\em Future work.}} All of these results reinforce the impression that the notion of causal graph dynamics we have reached is both general and robust; and at the same time concrete enough so that non-trivial facts can be said about them, and useful models be instantiated. Similar specific-purpose models are in fact in use in a variety of contexts ranging from social networks to epidemiology \cite{KozmaBarrat,Fleury} or physics \cite{QuantumGraphity1}. Several of these, however, include an extra ingredient of non-determinism, probabilities, or even quantum theory, which often are key ingredients of the physical situation at hand --- whereas the setting of this paper has remained nicely deterministic for now. Hence, studying probabilistic or quantum versions of causal graph dynamics ought to be promising.\\
There are plenty positive theoretical results about cellular automata that need to be reevaluated in this more general context, in terms of set-theoretical properties, structure, order, dynamical properties, computability etc. More interestingly even, negative results, such as the suspected anisotropy of Cellular automata, need to be reevaluated. Good wills are welcome. More concretely, we leave it as an open question whether condition $(ii)$ of Definition \ref{def:dynamics} can be relaxed to its binary form and the same results be obtained.

\section*{Acknowledgement} This research was funded by ANR CausaQ. We thank Rachid Echahed, Renan Fargetton, Miguel Lezama, Mehdi Mhalla, Simon Perdrix, Guillaume Theyssier, Eric Thierry and Nicolas Trotignon for inspiring discussions and pointers.\\
{\em This work is dedicated to Angelo.}

\end{document}